\documentclass[useAMS,usenatbib]{mn2e} 
\usepackage{amsmath}
\usepackage{graphicx} 
\usepackage{aas_macros}
\usepackage[usenames]{color} 
\usepackage{booktabs}
\usepackage{url}
\def\rmxaa{Revista Mexicana de Astronom{\'{\i}}a y Astrof{\'{\i}}sica}

\title[Pulsations of the $\delta$ Scuti Star AN Lyncis]{Pulsations and Period Variations of the $\delta$ Scuti Star \\AN Lyncis in a possible three-body system$^*$}
\author[]{Li, Gang$^{1}$; Fu, Jianning$^1$ $^{**}$; Su, Jie$^1$; Lester Fox-Machado$^2$; Raul Michel$^2$;
\newauthor{Guo, Zhen$^3$; Liu, Jinzhong$^4$; Feng, Guojie$^4$}\\
$^1$Department of Astronomy, Beijing Normal University, Beijing 100875, China\\
$^2$Observatorio Astr\'onomico Nacional, Institute de Astronom\'ia, Universidad Nacional Aut\'onoma de M\'exico, \\
$^{~}$AP.877, Ensenada 2860, M\'exico\\
$^3$Department of Astronomy, School of Physics, Peking University, Beijing 100871, China\\
$^4$Xinjiang Astronomical Observatory, Chinese Academy of Sciences, Urumqi 830011, China\\
$^*$Data collected from Xinglong station of National Astronomical Observatories of China, San Pedro Martir (SPM) \\$~~$Observatory of Mexico and Xinjiang Astronomical Observatory of China\\
$^{**}$Send offprint request to: jnfu@bnu.edu.cn}

\date{}

\def\LaTeX{L\kern-.36em\raise.3ex\hbox{a}\kern-.15em 
 T\kern-.1667em\lower.7ex\hbox{E}\kern-.125emX} 
 
\begin{document}

\maketitle 
 
\begin{abstract}\label{abstract}
Observations for the $\delta$ Scuti star AN Lyn have been made between 2008 and 2016 with the 85-cm telescope at Xinglong station of National Astronomical Observatories of China, the 84-cm telescope at SPM Observatory of Mexico and the Nanshan One meter Wide field Telescope of Xinjiang Observatory of China. Data in $V$ in 50 nights and in $R$ in 34 nights are obtained in total. The bi-site observations from both Xinglong Station and SPM Observatory in 2014 are analyzed with Fourier Decomposition to detect pulsation frequencies. Two independent frequencies are resolved, including one non-radial mode. A number of stellar model tracks are constructed with the MESA code and the fit of frequencies leads to the best-fit model with the stellar mass of $M = 1.70 \pm 0.05~\mathrm{M_{\odot}}$ , the metallicity abundance of $Z = 0.020 \pm 0.001$, the age of $1.33 \pm 0.01$ billion years and the period change rate $1/P\cdot \mathrm{d}P/\mathrm{d}t =1.06 \times 10^{-9} ~\mathrm{yr^{-1}} $, locating the star at the evolutionary stage close to the terminal age main sequence (TAMS). The O-C diagram provides the period change rate of $1/P \cdot \mathrm{d}P /\mathrm{d}t =4.5(8)\times10^{-7}~\mathrm{yr^{-1}}$. However, the period change rate calculated from the models is smaller in two orders than the one derived from the O-C diagram. Together with the sinusoidal function signature, the period variations are regarded to be dominated by the light-travel time effect of the orbital motion of a three-body system with two low-luminosity components, rather than the stellar evolutionary effect.
\end{abstract} 
 
\begin{keywords} 
stars: individual: AN Lyn - stars: variables: $\delta$ Scuti
\end{keywords}

\begin{figure}
\centering
\includegraphics[width=0.5\textwidth]{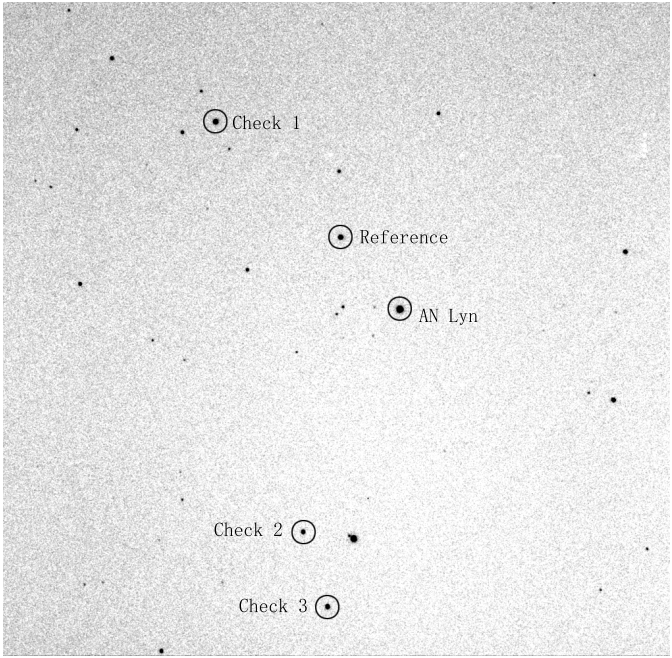}
\caption{A CCD image ($16.5'\times16.5'$) of AN Lyn, taken with the 85-cm telescope at Xinglong station of National Astronomical Observatories of China on March 13 of 2014. Reference marks the comparison star. Three check stars are pointed out as well. \label{fig:field_of_view}}
\end{figure}

\begin{figure}
\centering
\includegraphics[width=0.5\textwidth]{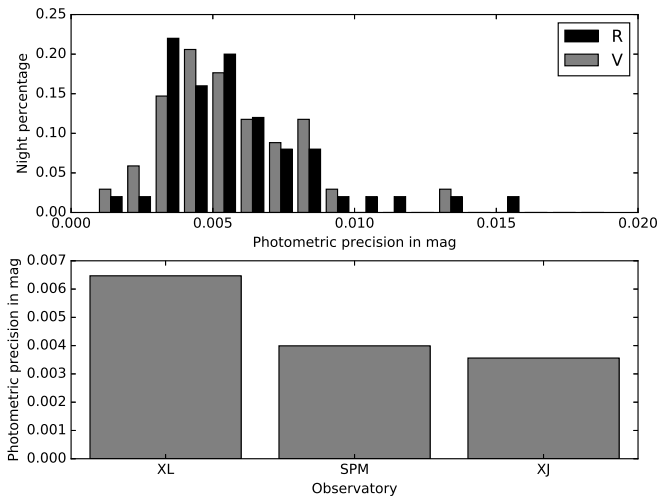}
\caption{Top: distribution of photometric precision of each run. One can find that more than 90\% data's precisions are better than $0^\mathrm{m}.01$. Bottom: mean photometric precision for all the runs of each telescope. XL stands for the 85-cm telescope in Xinglong station of National Astronomical Observatories of China. SPM stands for the 84-cm telescope at San P\'edro Martir (SPM) Observatory of Mexico. XJ is the Nanshan One meter Wide field Telescope of Xinjiang Astronomical Observatory of China. }
\label{fig:precision_distribution}
\end{figure}

\begin{table}
\caption{Observation journals of AN Lyn. XL85 means the 85-cm telescope at Xinglong station of China, SPM84 is the 84-cm telescope at SPM Observatory of Mexico and NOWT stands for the Nanshan One meter Wide field Telescope at Xinjiang Observatory of China. Prec stands for the mean photometric precision in mag during each run. \label{tab:observations}}
\scriptsize
\begin{tabular}{rrrrrrr}
\hline\hline
Run &Year & Month & Nights & Telescope & Frames & Prec.\\
\hline
$V$ band \\
1a & 2008 & Jan & 3 & XL85 & 2446 & 0.005\\
2a & 2008 & Feb & 3& XL85 &1777 &0.01\\
3a & 2010 & Dec &3 & XL85 &1565 &0.006\\
4a & 2013 & Feb &1& XL85  &435 &0.003\\
5a & 2013 & Apr &1& XL85  &396 &0.003\\
6a & 2014 & Feb &2 & XL85 &556 &0.006\\
7a & 2014 & Mar &13& XL85 SPM84 &2835 &0.006\\
8a & 2014 & Apr &1 & XL85 &91 &0.007\\
9a & 2015 & Jan &6& XL85  &2053 &0.004\\
10a & 2015 & Dec &5 & XL85 &1043 &0.006\\
11a & 2016 & Jan &9 & XL85 NOWT &1823 &0.005\\
$R$ band\\
5b & 2013 & Apr & 2 & XL85  & 841  & 0.006\\
6b & 2014  & Feb &  2 & XL85  & 543  & 0.007\\
7b & 2014  & Mar & 13 & XL85 SPM84 & 2889  & 0.007\\
8b & 2014  & Apr & 1 & XL85  & 87  & 0.02\\
9b & 2015  & Jan & 6& XL85   & 1943 &  0.006\\
10b & 2015  & Apr & 3 & XL85  & 397  & 0.006\\
11b & 2016  & Jan & 7& XL85 NOWT  & 1462  & 0.005\\
\hline
\end{tabular}
\end{table}


\begin{table*}
\begin{center}
\caption{Results of frequency analysis of the light curves obtained with the observation campaign in March of 2014. Freq stands for the frequency while Ampl is the amplitude. S/N is the signal over noise ratio.\label{tab:fourier transform}}
\begin{tabular}{rrrrrrrrr}
\hline\hline
  & \multicolumn{3}{c}{$V$ band }& \multicolumn{3}{c}{$R$ band}\\

\cmidrule(lr){2-5} \cmidrule(lr){6-9}

&Freq ($\mathrm{d}^{-1}$) & Ampl (mag) & S/N & sig &Freq ($\mathrm{d}^{-1}$) & Ampl (mag) & S/N & sig\\ 
\hline
$f_0$&	10.1721(2)&	 0.0868(17) &	 166.2 & 599.1 & 10.1720(5) & 0.0681(78) & 106.1 & 553.2\\ 
$2f_0$&	20.351(3)&	 0.0046(16) &	 11.6 & 89.2 & 20.357(5) & 0.0042(18) & 9.5 & 58.4\\
$f_2$ &	28.311(2)&	 0.0031(14) &	 6.6 & 44.2 &28.32(1) & 0.0024(13) & 7.4 & 23.6\\
$f_2-f_0$ &	18.125(6) &	 0.0028(14) & 6.0 & 40.0 & 18.126(6) & 0.0023(17) & 5.0 & 24.8\\
$f_2+f_0$&	38.496(6)&	 0.0021(12) &	 5.8 & 24.3 & 38.85(1) & 0.0018(13) & 9.8 & 17.9\\ 
\hline 
\end{tabular}
\end{center}
\end{table*}

\section{Introduction}\label{sec:introduction}
$\delta$ Scuti stars are pulsating variables locating in the classical Cepheid instability strip, 
on the main sequence or evolving from the main sequence to the giant branch \citep{2000ASPC..210....3B, 2000BaltA...9..149B}. The general consensus shows that most (possibly all) $\delta$ Scuti stars evolve in the main-sequence or the immediate post-main-sequence stages \citep{1973A&A....23..221B, 1979PASP...91....5B, 1980ApJ...235..153B}.
The amplitudes of pulsations in $\delta$ scuti stars are from mmag up to tenths of a magnitude \citep[see, e.g.,][]{2013RAA....13.1181N, 2017MNRAS.467.3122N, 2015AJ....149...84Z}. Their pulsation frequencies from ground-based observations are believed to be greater than 
5 $\mathrm{d}^{-1}$ but smaller than 50 $\mathrm{d}^{-1}$ \citep{2015MNRAS.452.3073B}, 
which is a good criterion for the variable classification. 
$\delta$ Scuti pulsators are divided into three groups based on the amplitude. The medium amplitude pulsators with visual amplitude between $0.^{\mathrm{m}}1$ and $0.^{\mathrm{m}}3$ are usually multiperiodic and the percentage of them is less than 5\% \citep[see the classification in,][]{1994A&AS..106...21R,1997A&A...324..959R}. As one of the medium amplitude minority, AN Lyn is an interesting target of investigation for the pulsations and period variations.


AN Lyn ($\mathrm{\alpha_{2000}=09^h14^m28^s}$, $\mathrm{\delta_{2000}=42^\circ46'38''}$, 10$^{\mathrm{m}}$.58 in $V$, A7IV-V) is a multiperiodic $\delta$ Scuti star with the principal period of $\mathrm{0^{d}.09827}$ and the visual amplitude of $\mathrm{\sim 0^m.18}$ \citep{2002A&A...385..503Z}. It has been observed from 1980 \citep{1981PASP...93...77Y} to 2016 with a large amount of data accumulated. The distance was estimated as 529 parsec by \cite{2015RMxAA..51...51P} using the $uvby-\beta$ photometry. A more convincing result was $720\pm160$ parsec, reported by $Gaia$ satellite \citep{2016A&A...595A...1G, 2016A&A...595A...2G, 2017A&A...599A..50A}. 

\cite{1997A&A...324..959R} pointed out that AN Lyn is a nearly cold and evolved $\delta$ Scuti star with solar metal abundance based on $uvby-\beta$ photometry, in which the `nearly cold' means that AN Lyn is closer to the red edge of the instability strip than the blue edge. The last report of AN Lyn's stellar parameters is $\log g=3.8$ and $\log T_\mathrm{eff}=3.8$ \citep{2015RMxAA..51...51P}. With Fourier analysis, \cite{1997A&A...328..235R} detected three independent frequencies, $f_1=10.1756,f_2=18.1309$ and $f_3=9.5598$ $\mathrm{d^{-1}}$, who denied that AN Lyn was a monoperiodic pulsator by earlier works. However, it is noticed that \cite{2002A&A...385..503Z} confirmed the existence of $f_1$ and $f_2$ but not $f_3$ in his work.

As far as the O-C analysis, it was hard for \cite{1997A&A...328..235R} to explain the unusual behaviour of AN Lyn. Since then, the binary nature has been gradually deduced from further O-C analysis. The O-C diagram shows a long-term increasing trend plus a fluctuating pattern. The long-term increasing trend is explained as the period increase due to the stellar evolution effect. \cite{2005AJ....130.2876H} reported the period change rate $\mathrm{d}P/\mathrm{d}t$ as $7.9\times10^{-10}$ d d$^{-1}$. \cite{2010PASJ...62..987L} provided the measurement of $2.09\left(\pm0.51\right)\times10^{-11}$ d d$^{-1}$. \cite{2015RMxAA..51...51P} reported the latest period change rate of $1.54\times10^{-10}$ d d$^{-1}$. The dramatic inconsistency comes from the under-sampling of the complex shape of the O-C curve. The fluctuating pattern in the O-C diagram is believed to be caused by the binary motion with the orbital period of $\mathrm{\sim}$ 20 years. \cite{2005AJ....130.2876H} reported the radial velocities of 11.4 km/s in 2003 and 36.8 km/s in 2004, respectively. The difference between these two values is taken as a proof of the binary nature. At the same time, the amplitude change of $f_1$ was detected as sine function with nearly the same period of binary motion \citep{2002A&A...385..503Z, 2005AJ....130.2876H,2010PASJ...62..987L,2015RMxAA..51...51P}. However, the reason for the amplitude variations is still an open question. 

In order to study the pulsations and period changes of AN Lyn in detail and explore its evolutionary status, we made observations for AN Lyn between 2008 to 2016 and constructed theoretical models to constrain the stellar parameters. Section~\ref{section:obs} introduces the observations. In Section~\ref{section:freq_analysis}, frequency analysis is performed with Fourier Decomposition. Section~\ref{section:theoretical model} provides constraints from theoretical models and Section~\ref{section:O-C} gives the O-C analysis. Finally, Section~\ref{section:discussion} and Section~\ref{section:conclutions} present discussions and conclusions, respectively.

\section{Observations}\label{section:obs}

AN Lyn had been observed with the 85-cm telescope in Xinglong station of National Astronomical Observatories of China between January of 2008 and January of 2016. During this period, a bi-site observation campaign was made with the 85-cm telescope at Xinglong station in China and the 84-cm telescope at San P\'edro Martir (SPM) Observatory in Mexico during March 13 to 20 of 2014. AN Lyn was also monitored from January 21 to 24 of 2016 with the Nanshan One meter Wide field Telescope of Xinjiang Astronomical Observatory of China. In total, data have been collected for AN Lyn for 50 nights in $V$ band and 34 nights in $R$ band, respectively. 

The data are reduced with the standard procedure of CCD photometry. After bias and flat-field correction, aperture photometry is performed by using the DAOPHOT program of IRAF. Figure~\ref{fig:field_of_view} shows a CCD image of AN Lyn, where a reference star (GSC 02990-00001) and three check stars (TYC 2990-221-1, GSC 02990-00139, GSC 02990-00335) are marked. The light curves were then produced by computing the magnitude differences between AN Lyn and the reference star and verified by the three check stars. 

Table~\ref{tab:observations} lists the observation runs and the mean photometric precisions of each run. Figure~\ref{fig:precision_distribution} shows the detailed distribution of photometric precisions at each night and the differences among three telescopes. One can find that the precisions in more than 90\% observation nights are better than $0^\mathrm{m}.01$ and the typical photometric precision is around $\sim 0^\mathrm{m}.006$. There are 19 nights in which the precisions are better than $0^\mathrm{m}.004$. However, the precisions during four nights are worse than $0^\mathrm{m}.01$, in which the worst one is $0^\mathrm{m}.016$. The precisions differ among the three telescopes. The 84-cm telescope at San P\'edro Martir (SPM) Observatory and the Nanshan One-meter Wide field Telescope of Xinjiang Astronomical Observatory perform consistently that their mean photometric precision for all the runs is $0^\mathrm{m}.004$ whereas the 85-cm telescope in Xinglong station shows a little worse result with the mean photometric precision of $0^\mathrm{m}.006$. There is no apparent distinction in different bands. The photometry in $R$ shows a slightly larger error.

\begin{figure}
\includegraphics[width=0.5\textwidth]{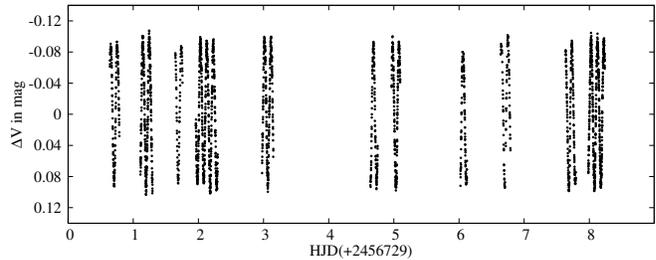}
\caption{Light curves of AN Lyn in $V$ during the bi-site observation campaign in March of 2014. \label{fig:light_curve_for_fourier}}
\end{figure}

\begin{figure}
\centering
\includegraphics[width=0.5\textwidth]{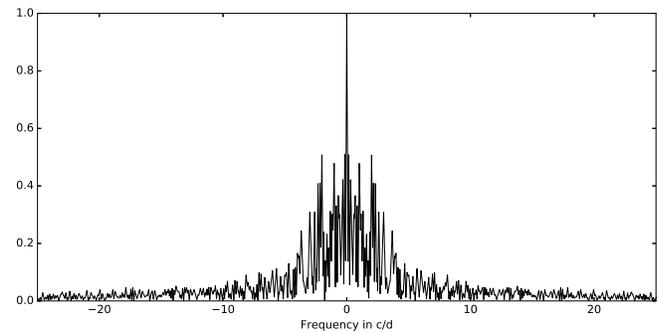}
\caption{Spectral window of the light curves in $V$ during the bi-site observation campaign in March of 2014.}\label{fig:spectral_window}
\end{figure}

\begin{figure*}
\centering
\includegraphics[width=1\textwidth]{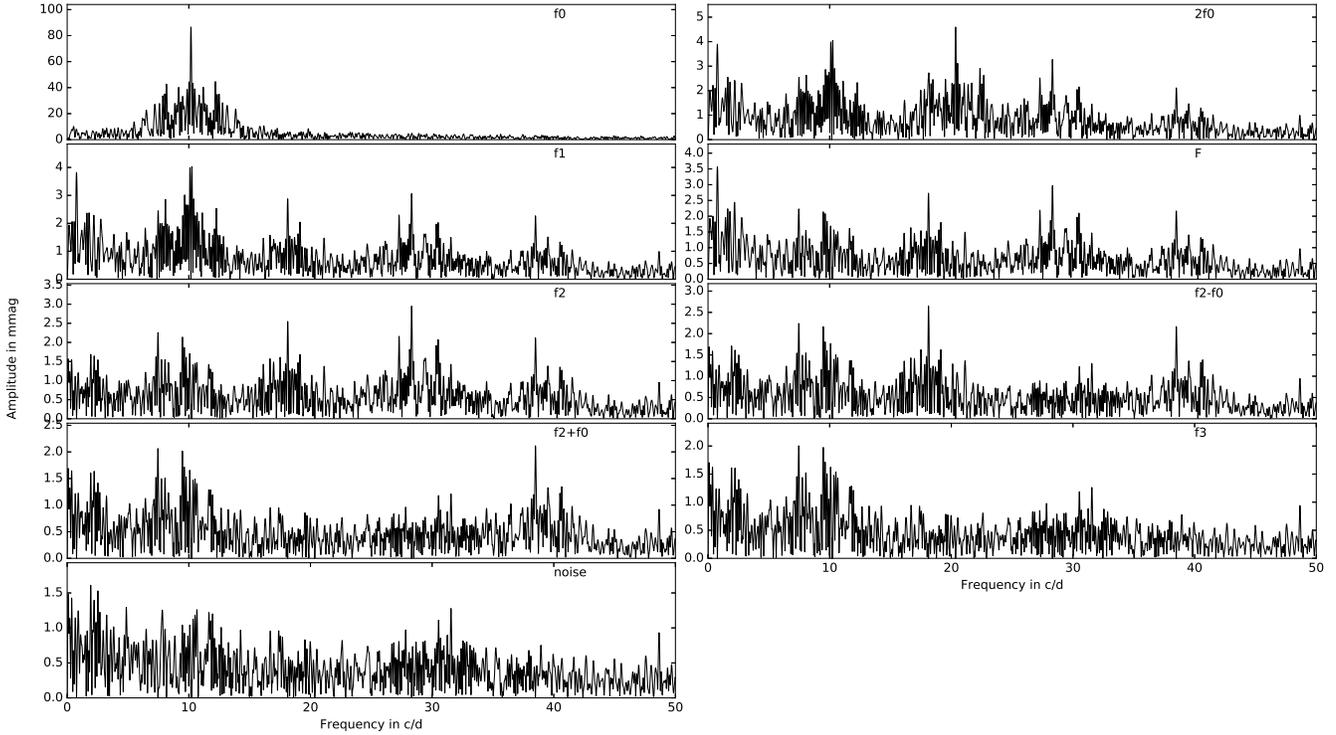}
\caption{The amplitude spectrum of the light curves in $V$ band from the bi-site observations in March of 2014 and the spectra of the light curves after prewhitenings of frequencies.}\label{fig:amp_spec_V}
\end{figure*}

\begin{figure*}
\centering
\includegraphics[width=1\textwidth]{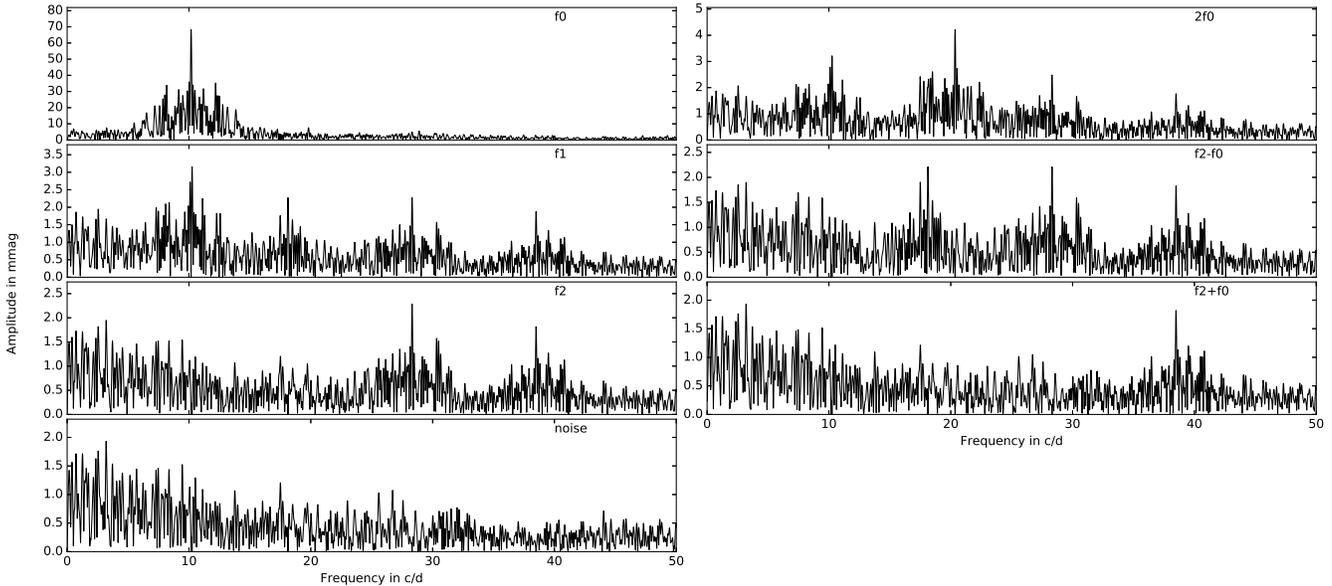}
\caption{Same as Figure \ref{fig:amp_spec_V} but in $R$ band.}\label{fig:amp_spec_R}
\end{figure*}
\begin{figure*}
\includegraphics[width=1\textwidth]{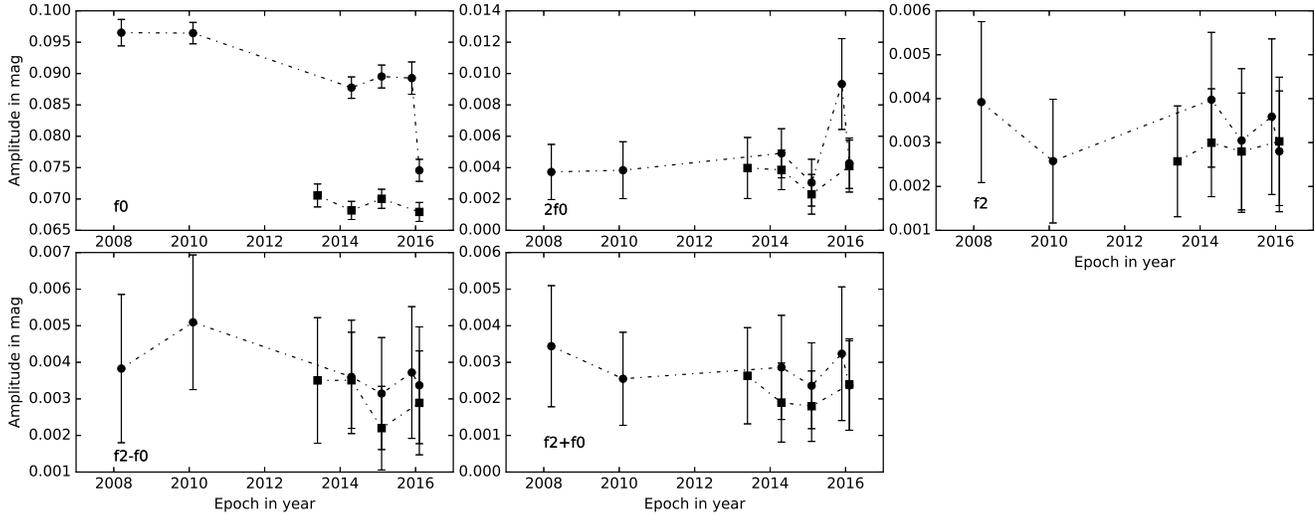}
\caption{Amplitude variations of five frequencies of AN Lyn listed in Table~\ref{tab:fourier transform}. The circle stands for the amplitude in $V$ band while the square is the amplitude in $R$ band. Note that the amplitudes of $f_0$ in $V$ band changed dramatically, especially in 2016.  \label{fig:amplitude_of_all_freq_variation}}
\end{figure*}

\begin{figure}
\includegraphics[width=0.5\textwidth]{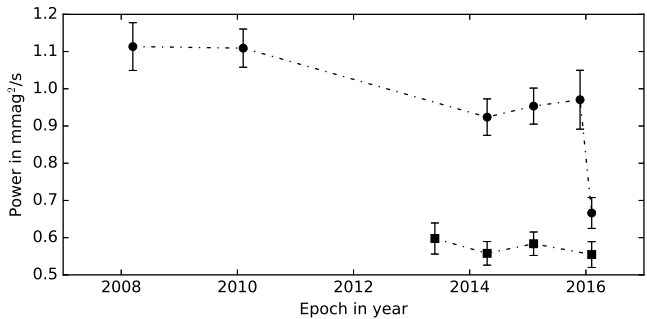}
\caption{The `pulsation power' of AN Lyn in different years. Dramatic variations are seen in 2016. \label{fig:magnitude_power}}
\end{figure}

\begin{table*}
\begin{center}
\caption{Amplitudes of the five frequencies of AN Lyn in $V$ in six combined observation datasets listed in Table~\ref{tab:observations}. Ampl stands for the amplitude of each frequency in mag in the six datasets, Error is the uncertainty of the amplitude in mag. The run names are from the first column of Table~\ref{tab:observations}.\label{tab:amplitude variation_V}}
\begin{tabular}{rrrrrrrrrrrrr}
\hline\hline
 & Ampl & Error & Ampl & Error & Ampl & Error & Ampl & Error & Ampl & Error & Ampl & Error \\
\cmidrule(lr){2-3} \cmidrule(lr){4-5} \cmidrule(lr){6-7} \cmidrule(lr){8-9} \cmidrule(lr){10-11}\cmidrule(lr){12-13}
Run & \multicolumn{2}{c}{1a+2a }& \multicolumn{2}{c}{3a}& \multicolumn{2}{c}{ 7a }& \multicolumn{2}{c}{9a}& \multicolumn{2}{c}{10a}& \multicolumn{2}{c}{11a}\\
\hline
$f_0$ & 0.0965 & 0.0021 & 0.0964 & 0.0017 & 0.0878 & 0.0017 & 0.0895 & 0.0018 & 0.0893 & 0.0026 & 0.0746 & 0.0018\\ 
$2f_0$ & 0.0037 & 0.0018 & 0.0038 & 0.0018 & 0.0049 & 0.0016 & 0.0030 & 0.0015 & 0.0093 & 0.0029 & 0.0043 & 0.0016\\ 
$f_2$ & 0.0039 & 0.0018 & 0.0026 & 0.0014 & 0.0040 & 0.0015 & 0.0030 & 0.0016 & 0.0036 & 0.0018 & 0.0028 & 0.0014\\ 
$f_2-f_0$ & 0.0038 & 0.0020 & 0.0051 & 0.0018 & 0.0036 & 0.0015 & 0.0031 & 0.0015 & 0.0037 & 0.0018 & 0.0034 & 0.0016\\ 
$f_2+f_0$ & 0.0034 & 0.0017 & 0.0025 & 0.0013 & 0.0029 & 0.0014 & 0.0024 & 0.0012 & 0.0032 & 0.0018 & 0.0024 & 0.0012\\ 
\hline
\end{tabular}
\end{center}
\end{table*}

\begin{table*}
\begin{center}
\caption{Same as Table~\ref{tab:amplitude variation_V} but in $R$ band. \label{tab:amplitude variation_R}}
\begin{tabular}{rrrrrrrrr}
\hline\hline
 & Ampl & Error & Ampl & Error & Ampl & Error & Ampl & Error \\
\cmidrule(lr){2-3} \cmidrule(lr){4-5} \cmidrule(lr){6-7} \cmidrule(lr){8-9} 
Run & \multicolumn{2}{c}{ 5b }& \multicolumn{2}{c}{ 7b }& \multicolumn{2}{c}{9b}& \multicolumn{2}{c}{11b}\\
\hline
$f_0$ & 0.0706 & 0.0018 & 0.0682 & 0.0014 & 0.0700 & 0.0015 & 0.0679 & 0.0015\\ 
$2f_0$ & 0.0040 & 0.0019 & 0.0039 & 0.0013 & 0.0023 & 0.0013 & 0.0041 & 0.0017\\ 
$f_2$ & 0.0026 & 0.0013 & 0.0030 & 0.0012 & 0.0028 & 0.0013 & 0.0030 & 0.0015\\ 
$f_2-f_0$ & 0.0035 & 0.0017 & 0.0035 & 0.0013 & 0.0022 & 0.0011 & 0.0029 & 0.0014\\ 
$f_2+f_0$ & 0.0026 & 0.0013 & 0.0019 & 0.0011 & 0.0018 & 0.0010 & 0.0024 & 0.0012\\
\hline
\end{tabular}
\end{center}
\end{table*}

\section{Pulsation properties}\label{section:freq_analysis}
\subsection{Frequency Analysis}

In this subsection, we report the frequency analysis of AN Lyn. The frequencies in $V$ and $R$ band are resolved using the \emph{Period04} package \citep{2004IAUS..224..786L} in parallel with the $Sigspec$ package \citep{2007A&A...467.1353R}. To be convincing enough, we accept the frequencies which must appear in two bands from these two independent packages at the same time. Finally, Two independent frequencies, $f_0=10.172(2)$, $f_2=28.311(2)$ c/d, as well as their harmonics and combinations $2f_0$, $f_2-f_0$, $f_2+f_0$, are resolved and confirmed.

Firstly, the data of the bi-site observation campaign for AN Lyn in $V$ in 2014 are used to perform Fourier Decomposition with the software Period04 to extract frequencies of pulsation. We correct the zero-points of the light curves in each night. The zero-points are set to zero by subtracting the mean value of the averaged maximum and minimum magnitudes in each night. Figure~\ref{fig:light_curve_for_fourier} shows the light curves of AN Lyn in $V$ during the bi-site observation campaign in March of 2014. Figure~\ref{fig:spectral_window} plots the spectral window and Figure~\ref{fig:amp_spec_V} depicts the power spectra of the light curves and those after prewhitenings of frequencies. The frequencies with signal-noise ratio (S/N) higher than 4 are taken following the criterion of \cite{1993A&A...271..482B}. 
The light curves are then fitted with the following formula,
\begin{equation}
m=m_0+\sum a_i \sin\left[2\pi\left(f_i t + \phi_i\right)\right]
\end{equation}
where $a_i$ is the amplitude, $f_i$ the frequency, $\phi_i$ the corresponding phase. 

Table~\ref{tab:fourier transform} lists the results of frequency analysis while Figure~\ref{fig:V_bi-site} depicts the fit curves and observed curves in detail. Eight frequencies are resolved with four independent ones identified: $f_0=10.1721\pm0.0005$, $f_1=10.251\pm0.006$, $f_2=28.311\pm0.003$, and $f_3=7.46\pm0.01~\mathrm{d^{-1}}$. 
$f_0$ is the most powerful signal, corresponding to the fundamental mode of AN Lyn with $l=0$ and $n=0$. The existence of $f_1$ causes `beat' phenomenon in light curves since its value is very close to $f_0$. However, the frequency of $f_2=28.311\pm0.003~\mathrm{d^{-1}}$ should be an independent frequency with higher amplitude and larger S/N than those of the frequency of $18.125 \pm 0.003~\mathrm{d^{-1}}$ mentioned in previous works. The latter should be the linear combination of $f_2-f_0$. Considering the frequency ratio of the fundamental to first overtone mode of 0.778 \citep{2000ASPC..210....3B}, the first overtone mode is not detected. The other three frequencies, $f_1$, $f_2$, $f_3$,  should belong to nonradial modes of AN Lyn. As the frequency $F=0.77\pm0.01~\mathrm{d^{-1}}$ is located in the low frequency region of $0-5~\mathrm{d^{-1}}$, we do not take it as an intrinsic frequency of pulsation of AN Lyn. However, $F$ is believed to be caused by the contamination from the variations of either the atmospheric transparency or the instability of the sensitivity of the detector during the nights.

Secondly, the data in $R$ band are collected simultaneously and Fourier Decomposition is also performed after the zero-point correction. The power spectrum is shown in Figure~\ref{fig:amp_spec_R} and the resolved frequencies are listed in Table~\ref{tab:fourier transform}. Figure~\ref{fig:R_bi-site} shows the fit curves and the observed data. In order to avoid meaningless low frequencies, we scan the frequency in the range of 5 c/d to 50 c/d \citep{2015MNRAS.452.3073B}.  We find $f_0$, $2f_0$, $f_1$, $f_2$, $f_2-f_0$, $f_2+f_0$ from the Fourier Decomposition in $R$ band while $f_3$ doesn't emerge, which is in general consistent with the results in $V$ band. The disappeared frequency $f_3$ is worth suspecting. Considering both the absence in $R$ and the relatively low S/N in $V$ with the value of 4.1, which is close to the lower limit of Breger's criterion, this frequency hence isn't considered as an real frequency.

Finally, \emph{Sigspec} is implemented to the data in $V$ and $R$ in order to check the frequencies extracted from previous processes. This package is based on an analytical solution of the probability that a Discrete Fourier Transform (DFT) peak of a given amplitude does not arise from white noise in a non-equally spaced data set \citep{2007A&A...467.1353R}. The statistical estimator called spectral significance ($sig$) are used to evaluate the reliabilities of peaks. $Sigspec$ derives the frequencies by iteratively prewhitening the most significant frequency until the estimator `$sig$' is less than 5. We find that the frequencies extracted by period04 show relative large significances hence they emerged at the first several frequencies from Sigspec. However, the frequency $f_1=10.251$ d$^{-1}$ doesn't appear from the Sigspec analysis. Considering that $f_1-f_0$ is smaller than the frequency resolution $1/T$ where $T$ is the observation duration, we don't accept $f_1$ as a convincing frequency. Several frequencies with $sig>5$ are extracted from \emph{Sigspec}. However, they are not be accepted because of their absence in the results of the Period04 analysis.

Although there is not any apparent photometric precision difference between $V$ and $R$, one can find that the frequencies resolved from the data in $V$ in general show higher signal to noise ratios and lower errors than those from the data in $R$. Hence the former is used in the following analysis. In short, based on the data in $V$ and $R$ and the independent results from Period04 and Sigspec, five pulsation frequencies are resolved and two frequencies are identified as independent frequencies, which are $f_0=10.1721\pm0.0005$, $f_2=28.311\pm0.003$ $\mathrm{d^{-1}}$, respectively.

\subsection{Amplitude Variations}\label{subsection: amplitude Measurement}
In order to study the amplitude variations from 8-year observations, 5 frequencies listed in Table~\ref{tab:fourier transform} in $V$ are used to fit light curves of AN Lyn with six datasets presented in this paper. We use the frequencies and combinations listed in Table~\ref{tab:fourier transform} to fit the light curves. 
We select the runs with enough data and then the fitting is implemented in the runs respectively. 
The fitting curves and observed data are shown in Figure~\ref{fig:Light_curves_in_V_band} for the light curves in $V$ and in Figure~\ref{fig:Light_curves_in_R_band} for the light curves in $R$, respectively. Table~\ref{tab:amplitude variation_V} and Table~\ref{tab:amplitude variation_R} list the amplitudes with errors of the 5 frequencies in the six datasets.
Figure~\ref{fig:amplitude_of_all_freq_variation} shows the amplitude variations of the independent frequencies and their linear combinations. One can find that only the amplitude of the main frequency $f_0$ in $V$ decreased dramatically from $0^\mathrm{m}.97$ in 2008 to $0^\mathrm{m}.76$ in 2016 while the amplitudes of the other frequencies kept relatively stable. The variations of $f_0$'s amplitude follow the previous prediction from \cite{2010PASJ...62..987L}, which is mentioned in Section~\ref{subsec: ampl_varia}.

Following \cite{2013MNRAS.429.1585F}, the `pulsation power' of AN Lyn can be calculated in $\mathrm{mmag^2 \cdot s^{-1}}$ by 
\begin{equation}
E=\sum f_i \times A^2_i
\end{equation}
in which $f_i$ the resolved frequency including the independent one and the linear combination, $A_i$ the amplitude. Figure~\ref{fig:magnitude_power} plots the variations of `pulsation power', showing that AN Lyn's pulsation power in $V$ band had been changing from 2008 to 2016, synchronizing with the changes of the amplitude of the dominant frequency as shown in Figure~\ref{fig:amplitude_of_all_freq_variation}.

\begin{figure}
\includegraphics[width=0.5\textwidth]{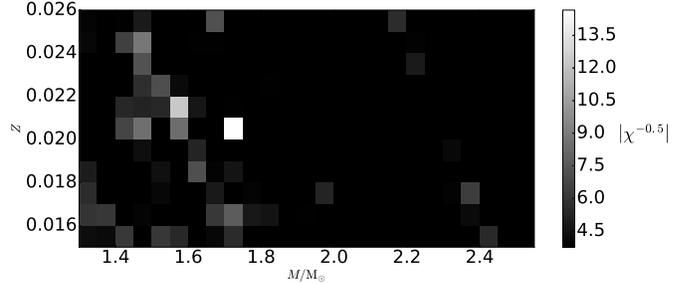}
\caption{Schematic relation among the value of $\chi$, the stellar mass $M$ and the metal abundance $Z$. The color stands for the value of $\left| \chi^{-0.5} \right|$. The lighter the color is, the better the calculated frequencies of the model fit the observed frequencies. \label{fig:stellar_model_M_Z}}
\end{figure}

\begin{figure}
\includegraphics[width=0.5\textwidth]{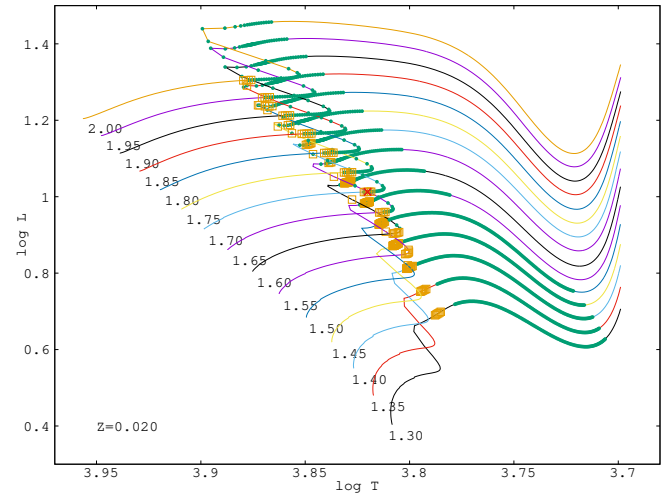}
\caption{Match between theoretical models and observations with $Z$ of 0.020. Solid lines show the evolutional tracks in different initial masses. The thick areas mean that the star satisfies $\log T \in \left[3.7,3.9\right]$ and $\log g \in \left[3.7,3.9\right]$. The squares show the models with $f_0 \in [10.0721,10.2721]$. The cross indicates the best-fit model with the minimum value of $\chi^2$. \label{fig:stellar_model}}
\end{figure}

\section{Constraints from theoretical models}\label{section:theoretical model}
We use Modules of Experiments in Stellar Astrophysics (MESA) to calculate stellar models. MESA is a group of source-open, powerful, efficient, thread-safe libraries for a wide range of applications in computational stellar astrophysics \citep{2011ApJS..192....3P,2013ApJS..208....4P}. MESA can simulate 1-D stellar evolution with a wide range of parameters from very-low mass to massive stars. 

The parameters we adjusted are the stellar mass $M$ and the metal abundances $Z$. The mass range is from $1.30\mathrm~{M_{\odot}}$ to $2.50\mathrm~{M_{\odot}}$ with the step of $0.05\mathrm~{M_{\odot}}$. The range of metal abundances is $Z\in[0.015,0.025]$ with the step of $0.001$. 275 evolutional tracks with different $M$ and $Z$ are calculated. 
The simulations are made from the pre-main-sequence to the moment when $\log T_{\mathrm{eff}}$ smaller than 3.7, where the star is after the end of post-main sequence. 
Then 275 evolutional tracks are calculated and the frequencies of the eigen modes are computed.
The frequencies in a small range of $f^{\mathrm{obs}}_i\pm0.1~\mathrm{d^{-1}}$ of different modes with degree $l \le 3$ are calculated, since the observable modes are believed to have $l \le 3$. The coefficients $n=0$ and $l=0$ are set for $f_0$ since it should be the fundamental frequency of AN Lyn. The best-fit model is determined by the minimum value of $\chi^2$
\begin{equation}
\chi^2=\frac{1}{2}\sum_{i=0}^1 \left( f^{\mathrm{cal}}_i-f^{\mathrm{obs}}_i\right)^2
\end{equation}
in which $f^{\mathrm{cal}}_i$ is the theoretically calculated frequency while $f^{\mathrm{obs}}_i$ the observed one. Figure~\ref{fig:stellar_model_M_Z} shows the relation among $\left| \chi^{-0.5} \right|$, $M$ and $Z$. Finally, a model with $M$ of $1.70~\mathrm{M_{\odot}}$ and $Z$ of $0.020$ is taken as the best-fit model due to the minimum $\chi^2$ value. 

Figure~\ref{fig:stellar_model} shows the theoretical evolutionary tracks where the best-fit model is marked with the cross on the H-R diagram. The best-fit model has the parameters as: $Z = 0.020 \pm 0.001$, $M = 1.70 \pm 0.05~\mathrm{M_{\odot}}$, $\log T_{\mathrm{eff}} = 3.8201 \pm 0.0002$, $\log g = 3.890 \pm 0.005$, $\log L = 1.0126 \pm 0.0003$, age = $(1.33 \pm 0.01) \times 10^9 ~\mathrm{yr}$ , $1/P\cdot \mathrm{d}P/\mathrm{d}t =
2.142 \times 10^{-9} ~\mathrm{yr^{-1}} $ with $\chi^2$ of $0.00005 ~\mathrm{d^{−2}}$. Table~\ref{tab:frequencies from MESA} lists the observed and corresponding theoretical frequencies of the best-fit model. The typical difference between the theoretical and the observed frequency is $0.001~\mathrm{d^{-1}}$.

\begin{figure*}
\includegraphics[width=0.8\textwidth]{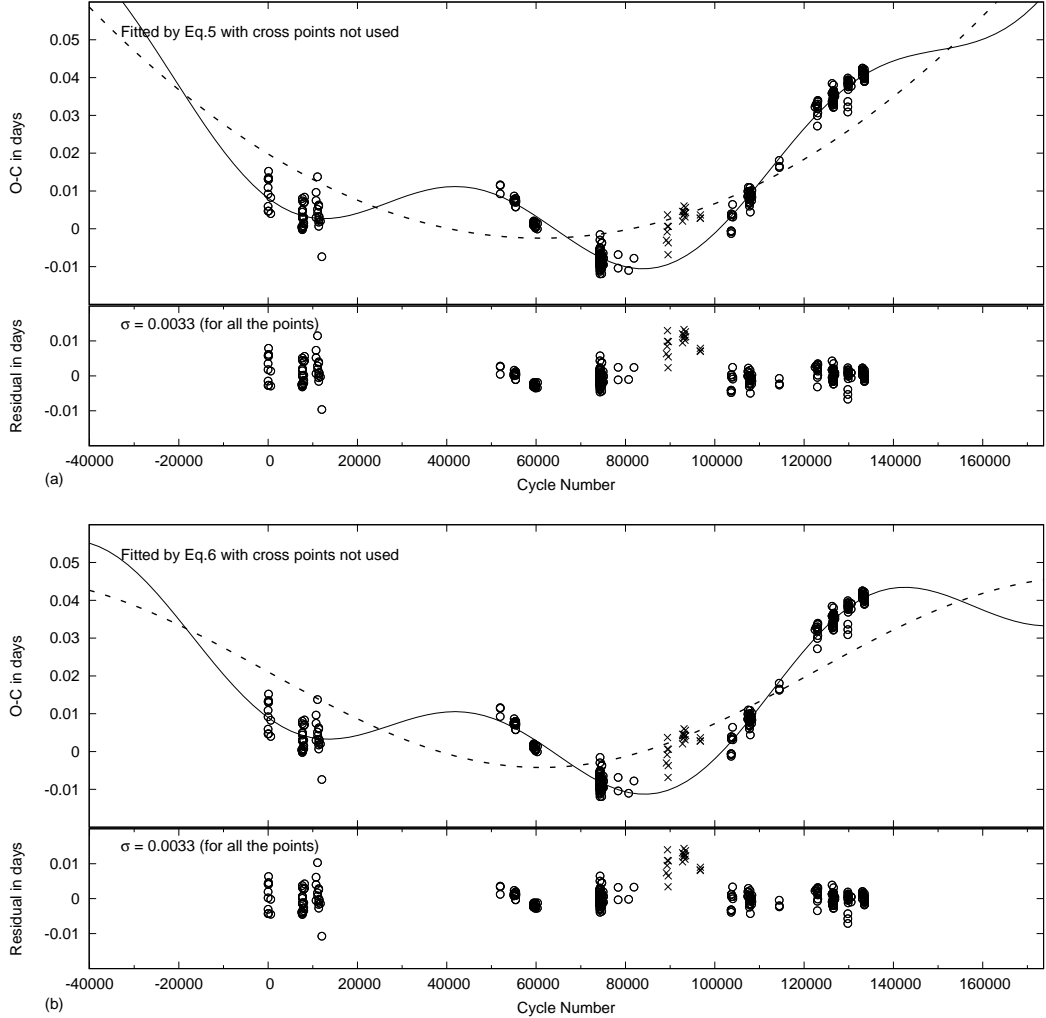}
\caption{O-C diagram of AN Lyn. (a): fitted by Equation~\ref{equ: porabolic and sinusoidal}. The dashed line is the parabola function while the solid line concerns both the parabola and the sine function. (b): fitted by Equation~\ref{equ: two_orbital_period_equation}. The dashed line shows the trend of long-period orbital motion while the solid line plots the whole motion caused by the two components. \label{fig:O_minus_C}}
\end{figure*}

\begin{table}
\begin{center}
\caption{Frequencies observed and calculated with the MESA code for the best-fit model of AN Lyn. $\delta f = f^{\mathrm{cal}}-f^{\mathrm{obs}}(\mathrm{d}^{-1})$.}\label{tab:frequencies from MESA}
\begin{tabular}{rrrrrr}
\hline\hline
& $f^{\mathrm{obs}}(\mathrm{d}^{-1})$ & $f^{\mathrm{cal}}(\mathrm{d}^{-1})$ & $n$ & $l$ & $\delta f$ ($d^{-1}$)\\ 
\hline
$f_0$&	10.1721&	 10.1727 &	 0 & 0 & 0.0006 \\
$f_2$ &	28.311 &	 28.304 & 7 & 1 & -0.007 \\
\hline 
\end{tabular}
\end{center}
\end{table}

 \cite{2015RMxAA..51...51P} gave the parameters of AN Lyn as $\log g=3.8\pm0.1$ and $\log T_{\mathrm{eff}}=3.8\pm0.1$, which are consistent with the corresponding values of the best-fit model within the range of three times of the uncertainties. As for the period change rate, there is an apparent conflict between the theoretical and the observed value. One can find that the prediction from the model is $1/P\cdot \mathrm{d}P/\mathrm{d}t =
1.06 \times 10^{-9} ~\mathrm{yr^{-1}} $ while the measurement from the O-C diagram mentioned in next Section is $4.5(8)\times10^{-7}~\mathrm{yr^{-1}}$, much larger by a factor of 424. 

The distance can be calculated based on the predicted $\log L$. The result, 453 pc, is not consistent with the photometry estimation and the parallax measurement mentioned in section~\ref{sec:introduction}. The discussion is made in section~\ref{subsection:three_body} in which the reason will be attributed to the components in the three-body system.

\section{O-C Analysis}\label{section:O-C}
O-C stands for O[bserved] \emph{minus} C[alculated] and is a widely used method when investigating the cyclic phenomena \citep{2005ASPC..335....3S}. We collect 238 times of maximum light of AN Lyn from literature, including 127 ones from \cite{2005AJ....130.2876H}, 108 from \cite{2010PASJ...62..987L} and 3 from \cite{2015RMxAA..51...51P}. We derive 124 times of maximum light from our observations, in which 64 ones in $V$ band and 35 in $R$ band determined from the data with the 85-cm telescope of Xinglong station of China, 9 in $V$ band and 9 in $R$ band with the 84-cm telescope of the SPM observatory of Mexico, 4 in $V$ band and 3 in $R$ band with the Nanshan One meter Wide field Telescope of Xinjiang Observatory of China. The times of maximum light are determined with the three-order polynomial fitting of the light curves. The errors are estimated with the Monte-Carlo simulation. Table~\ref{tab:O-C} lists the newly-determined times of maximum light. Figure~\ref{fig:O_minus_C} displays the O-C diagram. The ephemeris was calculated following \cite{2015RMxAA..51...51P}
\begin{equation}\label{equ:initial linear ephemeris}
HJD_\mathrm{{max}}=2444291.02077+0^\mathrm{d}.09827447\times E
\end{equation}
A parabolic fit is made for the O-C data in order to deduce the period variation rate ($1/P \cdot  \mathrm{d} P / \mathrm{d} t$). An undulatory shape of distribution of the residual emerges after removing the parabolic trend. Hence it is reasonable to fit the O-C data by including both the parabolic and the sinusoidal function as follows,
\begin{equation}　\label{equ: porabolic and sinusoidal}
O-C = \delta T_0 + \delta P E+\frac{1}{2}\frac{\mathrm{d} P}{\mathrm{d} t} P E^2 + A \sin \left[ 2\pi \left( \frac{E}{\Pi} + \phi \right) \right]　\\
\end{equation}
where $O-C$ means the value calculated by Equation~\ref{equ:initial linear ephemeris}. $\delta T_0$ means the correction of the first time of maximum light while $\delta P$ the correction of the pulsation period. $A$, $\Pi$ and $\phi$ stand for the amplitude, the period and the phase of the sinusoidal function, respectively. The fit is shown in Figure~\ref{fig:O_minus_C} (a). Three groups of points are not used in the fit because of their apparent deviations, marked with the black cross in Figure~\ref{fig:O_minus_C}. Table~\ref{tab:O-C poly fit} lists the coefficients of Equation~\ref{equ: porabolic and sinusoidal}. One can deduce that the period change rate $1/P \cdot \mathrm{d}P /\mathrm{d}t =4.5(8)\times10^{-7}~\mathrm{yr^{-1}}$ and the orbital period of binary is $23.0\pm0.3 $ yr.

\begin{table*}
\begin{center}
\caption{Coefficients of Equation~\ref{equ: porabolic and sinusoidal}. \label{tab:O-C poly fit}}
\begin{tabular}{rrrrrr}
\hline\hline
$\delta T_0$ (d) & $\delta P$ (d) & $P\mathrm{d}P/\mathrm{d}t$ (d) & $A$ (d) & $\Pi$ (d) & $\phi$ \\
\hline
$0.012(4)$ & $-7(2)\times10^{-7}$ & $1.2(2)\times10^{-11}$ & $0.0120(4)$ & $8.4(1)\times 10^3$ & $0.72(2)$\\
\hline

\end{tabular}
\end{center}
\end{table*}

Concerning the theoretical prediction from the model in Section~\ref{section:theoretical model}, the observation-determined period change rate is around 424 times larger than the theoretical value. Such a dramatic contradiction indicates an unusual cause of period change, other than the evolutionary effect. We guess that it could be caused by the light-travel time effect in a three-body system. The undulatory shape in the O-C diagram could be the evidence of a short-axis component (the second component, $M_2$) while the parabola trend reflects a long-axis massive component (the third component, $M_3$). We try to do Fourier Decomposition of the O-C data and fit it by using the two orbital frequencies,
\begin{equation}\label{equ: two_orbital_period_equation}
O-C=Z+\sum_{i=2,3} A_i \sin \left[ 2\pi \left( \frac{E}{\Pi_i} + \phi_i \right) \right] 
\end{equation}
in which $A_i$ is the amplitude of the orbital motion in days and $\Pi_i$ the period, $Z$ the zero-point. The result shows that there are two periods in the O-C data: $\Pi_2=24\pm 2 ~\mathrm{yr}$, $\Pi_3=66\pm 5 ~\mathrm{yr}$. The fit is shown in Figure~\ref{fig:O_minus_C} (b). 

The three groups of discarded data points marked with cross in Figure~\ref{fig:O_minus_C} show an unexpectable discrepancy with the fitted trend, which is the reason for them to be moved out in our analysis. \cite{2005IBVS.5643....1H,2005IBVS.5657....1H,2006IBVS.5731....1H,2006IBVS.5701....1K} reported the data and \cite{2010PASJ...62..987L} used them for O-C analysis firstly. However, by collecting more data, \cite{2015RMxAA..51...51P} presented a more convincing result, showing a systematic deviation of these three groups of data points from the fitted curve on the O-C diagram. Our observations provide six years of extension of data point distribution, supporting the fit of \cite{2015RMxAA..51...51P} rather than that of \cite{2010PASJ...62..987L}.

\section{Discussion}\label{section:discussion}

\subsection{Amplitude Variations}\label{subsec: ampl_varia}

Amplitude variations have been detected for almost all types of pulsating variable stars, such as the high-amplitude $\delta$ Scuti stars \citep{2002Ap&SS.281..699Z,2011AJ....142..100Z}, $\gamma$ Dor stars \citep{2013AAS...22135424R}, pulsating white dwarfs \citep{2013MNRAS.429.1585F,2003MNRAS.340.1031H,2015ApJ...809...14B} etc. The mechanisms that are responsible for the amplitude variations may be attributed to either intrinsic or external reasons. The latter includes the tidal effect due to a component star, like an interesting example given by \cite{2014MNRAS.437.1400B}. In this assumption, amplitude variations should be synchronized with the orbital motion.

\begin{figure}
\includegraphics[width=0.5\textwidth]{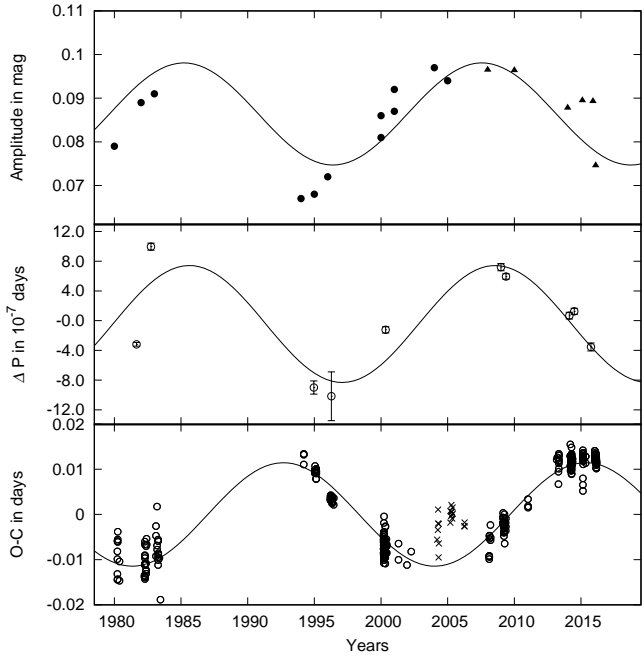}
\caption{Amplitude and period deviation of $f_0$ and O-C fluctuation. Top panel: Circle points stand for the amplitude data collected from \protect\cite{2010PASJ...62..987L} while the triangles the data from our observations. The period of amplitude variation is $22.4\pm1.1~\mathrm{yr}$. Middle panel: Period deviation of $f_0$ defined as $\Delta P=P-0^\mathrm{d}.09827447$ with periodicity of $22.9\pm1.5~\mathrm{yr}$. Bottom panel: O-C fluctuation with the parabola function extracted from Equation~\ref{equ: porabolic and sinusoidal} with the period of $23.0\pm 0.3~\mathrm{yr}$.} \label{fig:amplitude_period_O-C}
\end{figure}

We collect the amplitude and period variation data of $f_0$ from \cite{2010PASJ...62..987L} and added the data mentioned in Section \ref{subsection: amplitude Measurement}. The period in each epoch is calculated from the O-C's slopes extracted parabolic part. Figure~\ref{fig:amplitude_period_O-C} shows the amplitude, period deviation ($\Delta P=P-0^\mathrm{d}.09827447$) and O-C fluctuations of AN Lyn. 
The period of $f_0$ amplitude variation is $22.4\pm1.1~\mathrm{yr}$ and the one of $\Delta P$ is $22.9\pm1.5~\mathrm{yr}$, while the period of the O-C data is $23.0\pm 0.3~\mathrm{yr}$. The accordance of these three values of period is a strong indication that the companion star can change the pulsation amplitude of AN Lyn. 
The maximum value of orbital radial velocity is only $2.4~\mathrm{km/s}$ calculated with Doppler effect $\Delta P_{\mathrm{max}}/P=v/c$. Nevertheless, two radial velocities provided by \cite{2005AJ....130.2876H} varied very much (11.4 km/s in 2003 and 36.8 km/s in 2004), which might be due to the zero-point shift of the instrument, rather than the motion of AN Lyn due to binarity.

\subsection{Three-body Hypothesis}\label{subsection:three_body}

\begin{figure}
\includegraphics[width=0.5\textwidth]{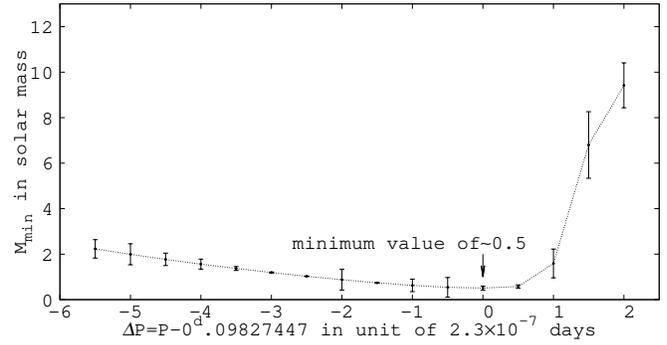}
\caption{Relation between the lower limit of the third component's mass and the initial period value in the ephemeris equation.}\label{fig:delta_P_to_m3}
\end{figure}

\begin{table}
\begin{center}
\caption{Parameters of the three-body system. Freq stands for the frequency of the O-C diagram; $P$ correspond period of frequency; Amp for the amplitude; $a_\mathrm{{min}}$ the lower limit of the semi-major axis while $M_\mathrm{{min}}$ the lower limit of the component's mass. \label{tab:O-C sin fit}}
\scriptsize
\begin{tabular}{rrrrrrr}
\hline\hline
 &Freq($\times 10^{-6}\mathrm{E}^{-1}$) & $P$(yr) & Amp(d) & $a_\mathrm{{min}}$(au) & $M_\mathrm{{min}}$ ($\mathrm{M_{\odot}}$) \\ 
\hline
$F_1$&	4.1(3) &	 66(5) & 0.025(3) & 4.3(5) & 0.51(8) \\
$F_2$ &	11(1) &	 24(2) & 0.012(3) & 2.1(5) & 0.4(1) \\
\hline 
\end{tabular}
\end{center}
\end{table}

As illustrated in Section \ref{section:theoretical model} and Section \ref{section:O-C}, the contradiction of period change rate from the O-C observations and the theoretical predictions implied the possibility of the existence of three components of the AN Lyn system. Based on the orbital period and the lower limit of semi-axis deduced by the O-C's amplitudes, the lower limit of the two components' masses can be estimated by the Kepler third law:
\begin{equation}
4\pi^2 \frac{a_2^3}{T_2^2}=\mathrm{G}\frac{m_2^3}{\left( m_1+m_2 \right)^2}
\end{equation}
\begin{equation}
4\pi^2 \frac{a_3^3}{T_3^2}=\mathrm{G}\frac{m_3^3}{\left( m_1+m_2+m_3 \right)^2}
\end{equation}
where $m_1$, $m_2$ and $m_3$ are the masses of AN Lyn and the two companions, $a_{2,3}$ and $T_{2,3}$ are the semi-axes and the periods of the two orbits, respectively. 
Due to the stability of this system, the axis of the orbit of the third component should be long enough so that the gravity force on the third component equals approximately to the synthesised force of gravity from both AN Lyn and the second component. To be quantitative, the ratio of $a_3$ and $a_2$ should be larger than or equal to 3 \citep{1972CeMec...6..322H}. 
The parameters of the three-body system are listed in Table~\ref{tab:O-C sin fit}. The existence of two components with minimum masses of $\sim 0.4 \mathrm{M_{\odot}}$ and $\sim 0.5 \mathrm{M_{\odot}}$ respectively implies that the evolutionary history of the AN Lyn system is mysterious. \cite{2005AJ....130.2876H} displayed a high resolution spectrum of AN Lyn in the region of H$\beta$. It seems that there isn't any spectral line from other stars, which indicates that the luminosities of the two components should be much lower than that of AN Lyn, implying that they are compact stars or red dwarf. 

However, the axis of the third component's orbit $a_3$ depends on the initial value of period in Equation~\ref{equ:initial linear ephemeris}. A wrong period value would enlarge variations in the O-C diagram, and increase $a_3$, making the parameters of the third component uncertain. We fit the O-C data calculated by different initial linear ephemerides and deduce the lower limit of the third component's mass displayed in Figure~\ref{fig:delta_P_to_m3}. One can find that the minimum value of mass of the third component is $\sim 0.5\mathrm{M_{\odot}}$, corresponding to the initial ephemeris in Equation~\ref{equ:initial linear ephemeris}.

The stellar model given in section~\ref{section:theoretical model} reveals the luminosity of AN Lyn and the distance can be calculated based on the luminosity and the visual magnitude. The result is 453 pc, which is narrowly consistent with $Gaia$'s observation within the range of three times of uncertainty. The huge discrepancy may be posed by the influence of the light emitted from other two components in the three-body system. The asteroseismic analysis only gives the luminosity of AN Lyn. However, the visual magnitude from the photometry is the sum of three components' brightnesses. The visual magnitude and the absolute magnitude are mismatched so that the system is brighter than that AN Lyn should be. Finally, the distance is shorter than that from the the parallax observation.

\section{Conclusions}\label{section:conclutions}

After 8-year observations, the $\delta$ Scuti star AN Lyn are investigated and analysed further. Two independent frequencies are resolved from a one-week bi-site observation campaign, including one non-radial mode. Amplitude variations of the three frequencies and their linear combinations are detected. The amplitude of the main frequency in $V$ has been changing with time while the other frequencies keep their amplitudes stable.

Stellar models are constructed and the theoretical frequencies are fitted with the two observed frequencies. The minimum value of $\chi^2$ gives the best-fit model, showing that AN Lyn locates near the terminal age main sequence on the H-R diagram with $M$ of $1.70\pm 0.05~\mathrm{M_{\odot}}$, $Z$ of $0.020\pm 0.001$ and age of $1.33 \pm 0.01$ billion years. 

O-C diagram of AN Lyn is made with the new observation data together with those in the literature. Since the determined period change rate is much larger than the theoretical prediction due to the evolutionary effect, the variations in the O-C diagram show the evidence of the light-travel time effect in a three-body system with two low-luminosity components, with masses larger than $0.4\mathrm{M_{\odot}}$ and $0.5\mathrm{M_{\odot}}$, respectively.

Future observation is needed to confirm this hypothesis, including both high-quality time-series photometry and radial velocity observations. Spectroscopic observations in the ultraviolet and infrared bands are also necessary in order to confirm the type of the components.

\section*{Acknowledgements} 

JNF acknowledges the support from both the NSFC grant 11673003 and the National Basic Research Program of China (973 Program 2014CB845700). JZL acknowledges the support from the CAS `Light of West China' program (2015-XBQN-A-02). LFM acknowledges the financial support from the UNAM under grant PAPIIT IN 105115. This work has made use of data from the European Space Agency (ESA) mission $Gaia$ (\url{https://www.cosmos.esa.int/gaia}), processed by the $Gaia$ Data Processing and Analysis Consortium (DPAC, \url{https://www.cosmos.esa.int/web/gaia/dpac/consortium}). Funding for the DPAC has been provided by national institutions, in particular the institutions participating in the $Gaia$ Multilateral Agreement.

\bibliographystyle{mn2e}
\bibliography{hotgdor}

\appendix

\section{Light curves and new maximum times of AN Lyn}
\label{appendix:light_curves}

\begin{figure*}
\includegraphics[width=1\textwidth]{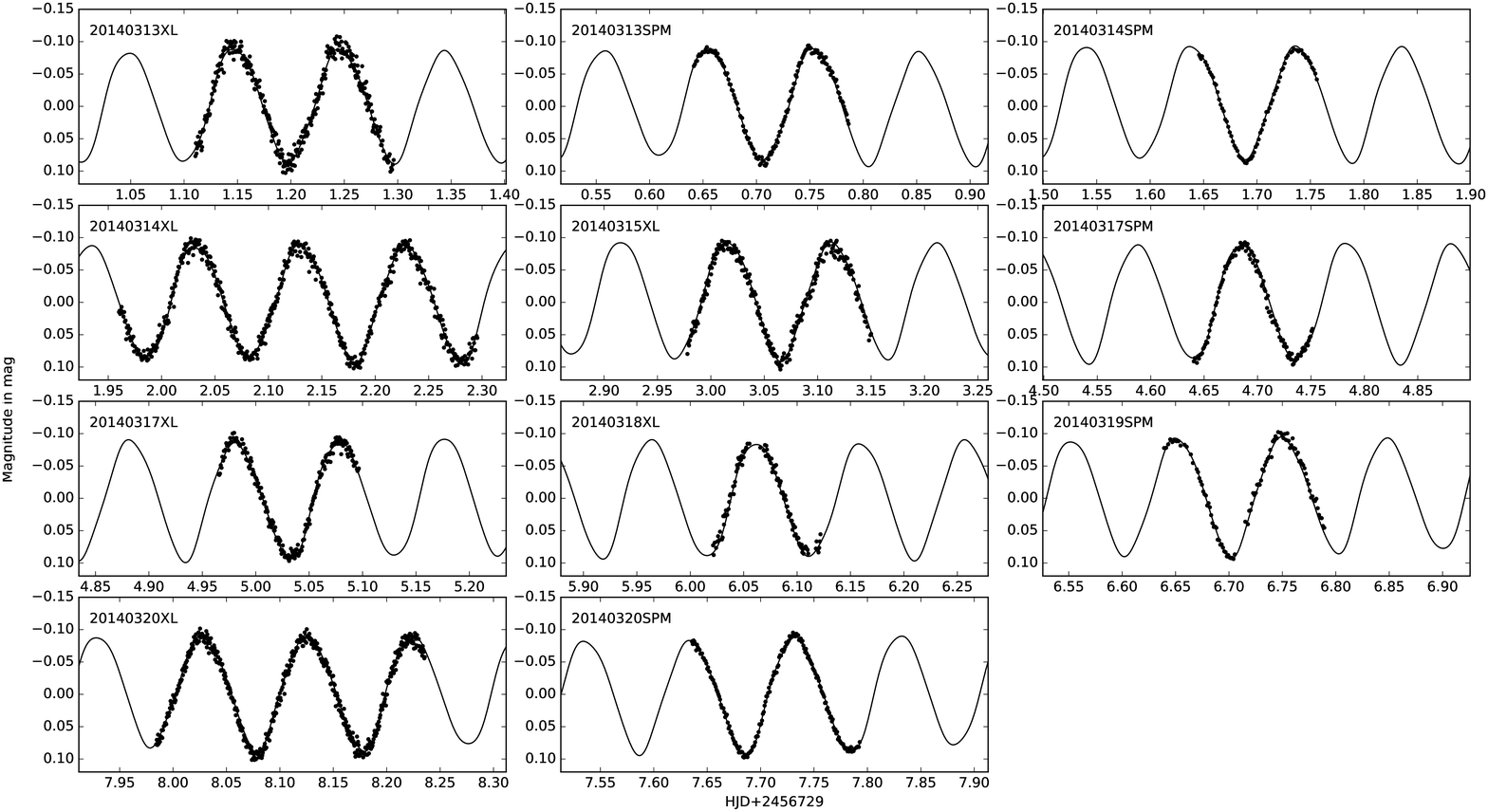}
\caption{Light curves and fitted results in $V$ during the bi-site observation campaign in March of 2014. The solid lines stand for the fitting curves with the five frequencies listed in Table~\ref{tab:fourier transform}. XL is the 85-cm telescope at Xinglong station of National Astronomical Observatories of China. SPM is the 84-cm telescope at San P\'edro Martir (SPM) Observatory of Mexico and XJ stands for the Nanshan One meter Wide field Telescope of Xinjiang Observatory of China. \label{fig:V_bi-site}}
\end{figure*}

\begin{figure*}
\includegraphics[width=1\textwidth]{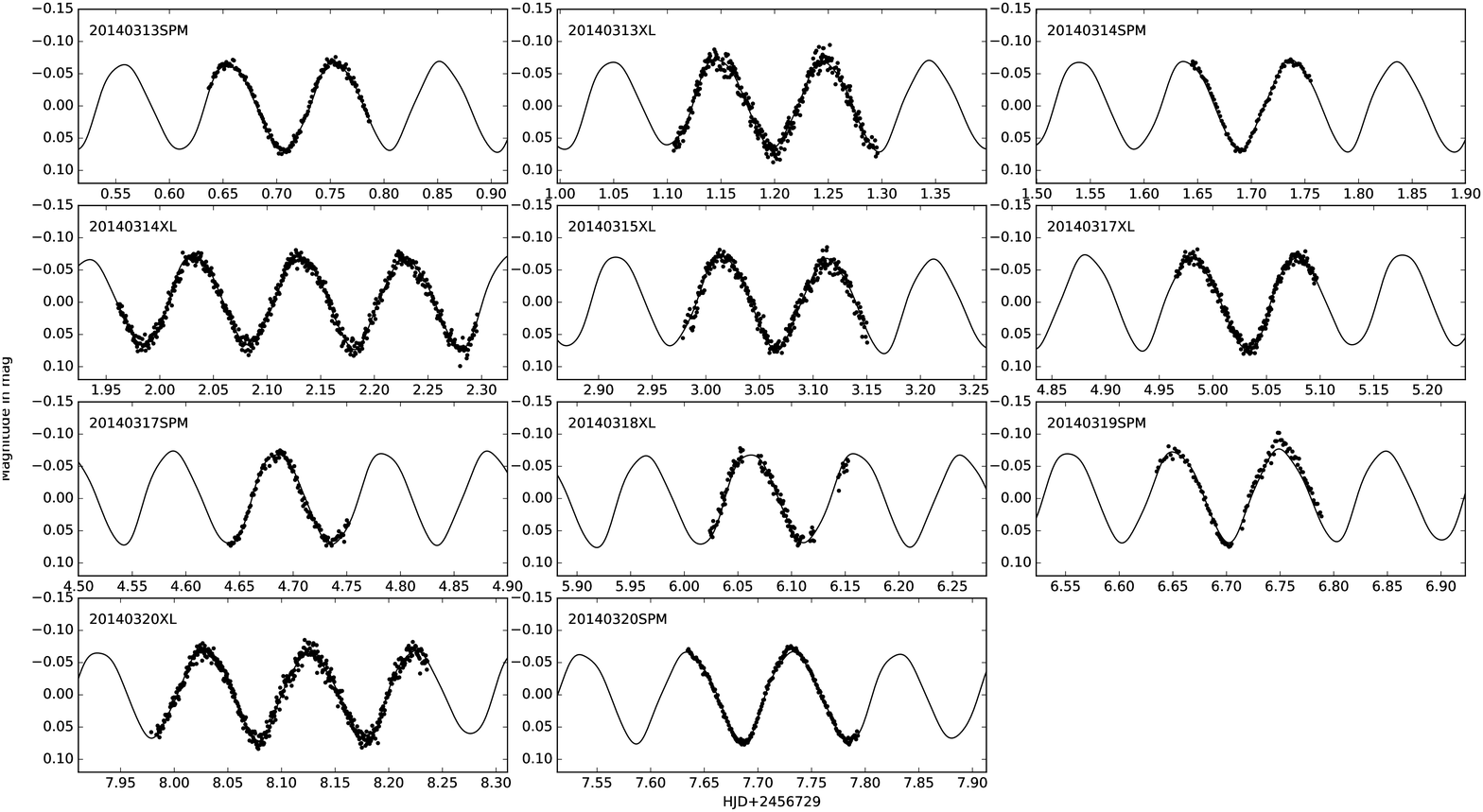}
\caption{Same as Figure~\ref{fig:V_bi-site} but in $R$ band. \label{fig:R_bi-site}}
\end{figure*}

\begin{figure*}
\includegraphics[width=0.95\textwidth]{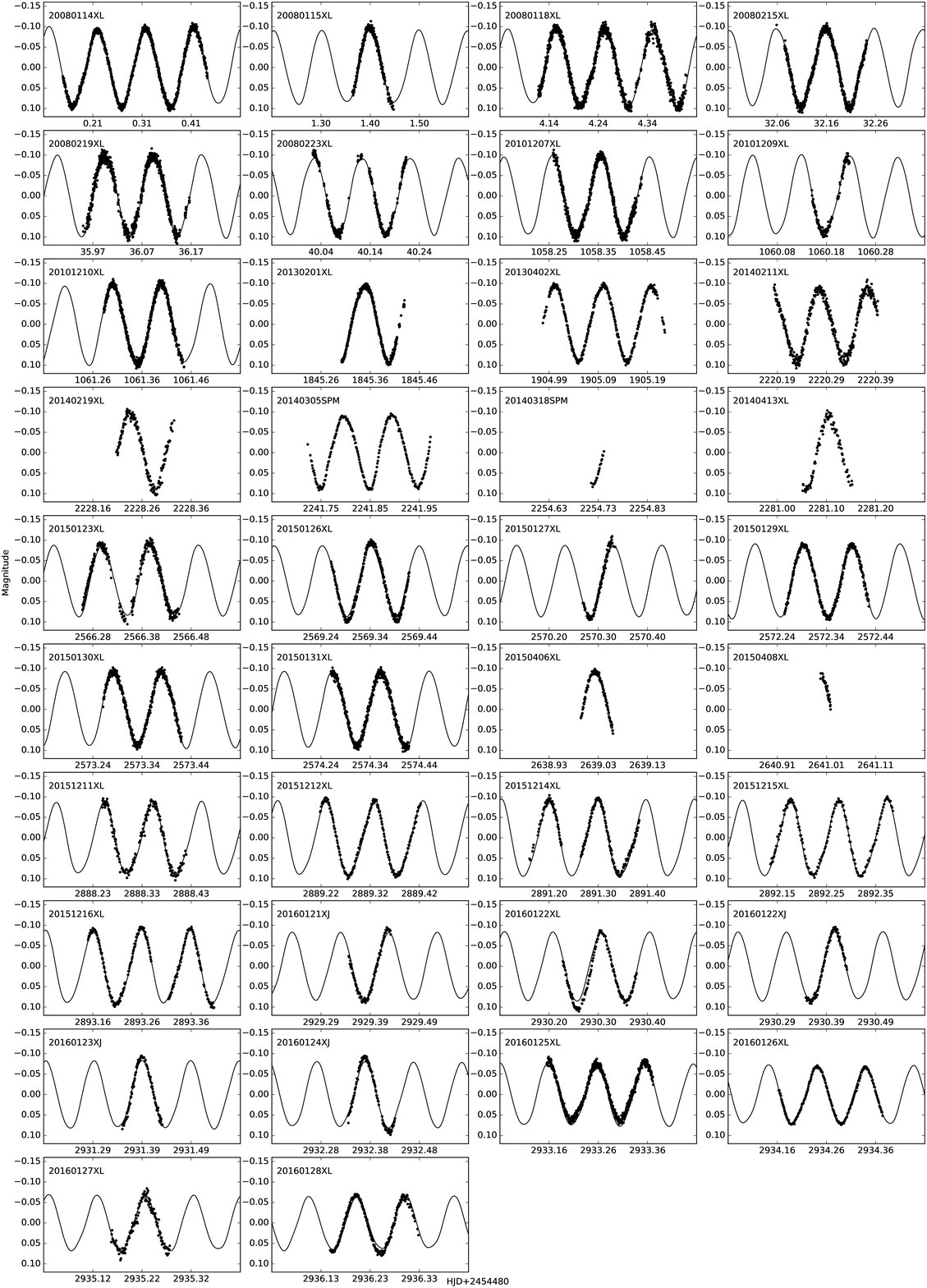}
\caption{All the light curves collected for AN Lyn in $V$ band. The solid lines stand for the fitting curves with the five frequencies listed in Table~\ref{tab:fourier transform}. \label{fig:Light_curves_in_V_band}}
\end{figure*}

\begin{figure*}
\includegraphics[width=1\textwidth]{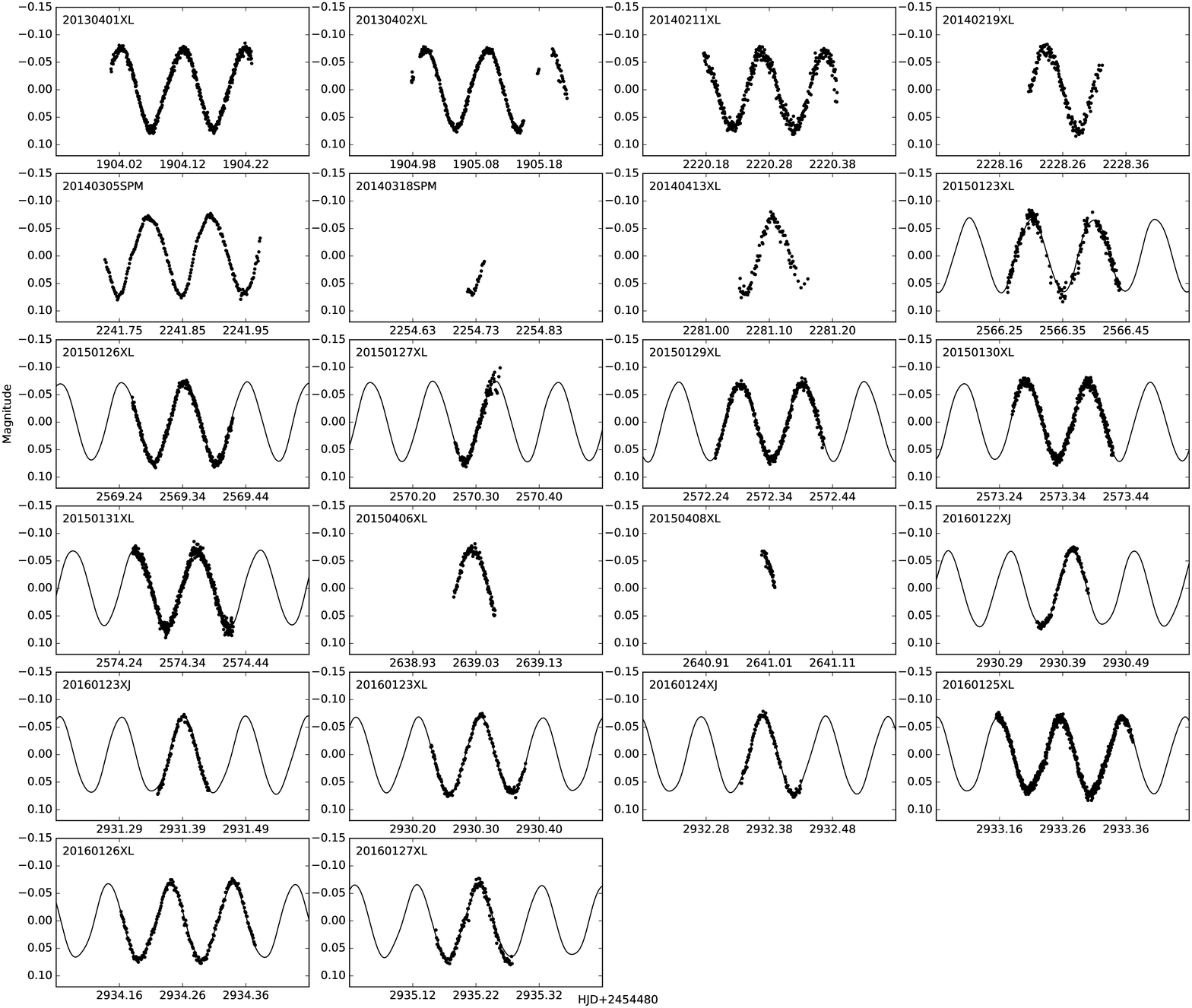}
\caption{Same as Figure~\ref{fig:Light_curves_in_V_band} but in $R$ band. \label{fig:Light_curves_in_R_band}}
\end{figure*}


\begin{table*}
\begin{center}
\scriptsize
\caption{Newly-determined times of maximum light of AN Lyn. $T_\mathrm{{max}}$ is the times of maximum light in Helio-centric Date-2450000. \label{tab:O-C}}
\begin{tabular}{rrrrrrrrrrrr}
\hline\hline
$T_\mathrm{{max}}$ & Error & Band &$T_\mathrm{{max}}$ & Error & Band &$T_\mathrm{{max}}$ & Error & Band &$T_\mathrm{{max}}$ & Error & Band   \\
\hline
4480.2155 & 0.0011 & $V$ & 4480.3131 & 0.0005 & $V$ & 4480.4121 & 0.0007 & $V$ & 4481.39461 & 0.00027 & $V$ \\
4484.1509 & 0.0010 & $V$ & 4484.2486 & 0.0010 & $V$ & 4484.3476 & 0.0006 & $V$ & 4512.16174 & 0.00005 & $V$ \\
4515.9911 & 0.0020 & $V$ & 4516.0900 & 0.0016 & $V$ & 5538.3554 & 0.0011 & $V$ & 5541.3018 & 0.0003 & $V$ \\
5541.4004 & 0.0003 & $V$ & 6325.3515 & 0.0004 & $V$ & 6384.0227 & 0.0015 & $R$ & 6384.1204 & 0.0016 & $R$ \\
6384.2182 & 0.0018 & $R$ & 6385.0037 & 0.0015 & $V$ & 6385.0042 & 0.0024 & $R$ & 6385.1038 & 0.0008 & $V$ \\
6385.1041 & 0.0010 & $R$ & 6385.1991 & 0.0010 & $V$ & 6700.2722 & 0.0014 & $R$ & 6700.2723 & 0.0012 & $V$ \\
6700.3683 & 0.0015 & $V$ & 6700.373 & 0.004 & $R$ & 6708.2293 & 0.0021 & $R$ & 6708.2303 & 0.0018 & $V$ \\
6721.7932 & 0.0027 & $V$ & 6721.7939 & 0.0020 & $R$ & 6721.8912 & 0.0025 & $V$ & 6721.8913 & 0.0017 & $R$ \\
6729.654 & 0.003 & $R$ & 6729.655 & 0.003 & $V$ & 6729.7519 & 0.0027 & $R$ & 6729.752 & 0.003 & $V$ \\
6730.1455 & 0.0016 & $V$ & 6730.1461 & 0.0016 & $R$ & 6730.2452 & 0.0017 & $R$ & 6730.246 & 0.003 & $V$ \\
6730.736 & 0.0021 & $V$ & 6730.738 & 0.003 & $R$ & 6731.0304 & 0.0014 & $V$ & 6731.0317 & 0.0009 & $R$ \\
6731.1275 & 0.0014 & $R$ & 6731.1285 & 0.0013 & $V$ & 6731.225 & 0.003 & $R$ & 6731.2268 & 0.0008 & $V$ \\
6732.0131 & 0.0012 & $V$ & 6732.014 & 0.004 & $R$ & 6732.1115 & 0.0016 & $V$ & 6732.113 & 0.005 & $R$ \\
6733.685 & 0.004 & $V$ & 6733.6877 & 0.0018 & $R$ & 6733.9796 & 0.0008 & $V$ & 6733.9800 & 0.0013 & $R$ \\
6734.0782 & 0.0010 & $R$ & 6734.0793 & 0.0006 & $V$ & 6735.058 & 0.005 & $R$ & 6735.059 & 0.004 & $V$ \\
6735.648 & 0.004 & $V$ & 6735.6513 & 0.0016 & $R$ & 6735.7495 & 0.0004 & $V$ & 6735.7501 & 0.0004 & $R$ \\
6736.7316 & 0.0005 & $R$ & 6736.7321 & 0.0005 & $V$ & 6737.0262 & 0.0010 & $V$ & 6737.0263 & 0.0008 & $R$ \\
6737.1243 & 0.0008 & $V$ & 6737.125 & 0.0005 & $R$ & 6737.2223 & 0.0020 & $V$ & 6737.2224 & 0.0004 & $R$ \\
6761.1033 & 0.0009 & $R$ & 6761.104 & 0.0005 & $V$ & 7046.2917 & 0.0008 & $V$ & 7046.2930 & 0.0004 & $V$ \\
7046.3928 & 0.0019 & $V$ & 7046.39897 & 0.00005 & $R$ & 7049.3454 & 0.0009 & $R$ & 7049.3458 & 0.0004 & $V$ \\
7052.2948 & 0.0014 & $R$ & 7052.2950 & 0.0004 & $V$ & 7052.3925 & 0.0008 & $V$ & 7052.3925 & 0.0008 & $R$ \\
7053.2751 & 0.0010 & $R$ & 7053.2767 & 0.0011 & $V$ & 7053.3741 & 0.0005 & $V$ & 7053.3756 & 0.0006 & $R$ \\
7054.357 & 0.0004 & $R$ & 7054.3574 & 0.0009 & $V$ & 7119.0230 & 0.0010 & $V$ & 7119.0231 & 0.0010 & $R$ \\
7120.1026 & 0.0013 & $R$ & 7368.3477 & 0.0006 & $V$ & 7369.2311 & 0.0017 & $V$ & 7369.3315 & 0.0011 & $V$ \\
7371.1975 & 0.0004 & $V$ & 7371.297 & 0.0003 & $V$ & 7372.181 & 0.0023 & $V$ & 7372.2787 & 0.0019 & $V$ \\
7372.3772 & 0.0022 & $V$ & 7373.1630 & 0.0019 & $V$ & 7373.2603 & 0.0008 & $V$ & 7373.3595 & 0.0011 & $V$ \\
7409.425 & 0.015 & $V$ & 7410.3105 & 0.0022 & $V$ & 7410.3105 & 0.0007 & $V$ & 7410.3109 & 0.0010 & $R$ \\
7410.4087 & 0.0012 & $V$ & 7410.4099 & 0.0018 & $R$ & 7411.3903 & 0.0020 & $R$ & 7411.3905 & 0.0008 & $V$ \\
7412.3722 & 0.0015 & $V$ & 7412.3728 & 0.0025 & $R$ & 7413.2578 & 0.0004 & $R$ & 7413.2584 & 0.0010 & $V$ \\
7413.3549 & 0.0010 & $V$ & 7413.3551 & 0.0009 & $R$ & 7414.2405 & 0.0009 & $V$ & 7414.2416 & 0.0017 & $R$ \\
7414.3387 & 0.0009 & $V$ & 7414.3390 & 0.0028 & $R$ & 7415.2235 & 0.0023 & $R$ & 7415.2236 & 0.0012 & $V$ \\
7416.2059 & 0.0004 & $V$ & 7416.3049 & 0.0006 & $V$ \\
\hline
\end{tabular}
\end{center}
\end{table*}

\label{lastpage}

\end{document}